\documentclass[
]{jpsj3}
\usepackage{txfonts}

\newcommand{\kab}{$\kappa_{ab}$}
\newcommand{\tc}{{\it T}$_{\rm c}$}
\newcommand{\degree}{$^{\circ}$}
\newcommand{\FeSeTe}{FeSe$_{1-x}$Te$_{x}$}
\newcommand{\Te}{FeSe$_{0.3}$Te$_{0.7}$}
\newcommand{\abpl}{{\it ab}-plane}
\newcommand{\kabHk}{$\kappa_{ab}$({\it H})/$\kappa_{ab}$(0)}
\newcommand{\kel}{$\kappa_{\rm el}(T)$}
\newcommand{\kph}{$\kappa_{\rm ph}(T)$}
\newcommand{\aT}{$\alpha(T)$}
\newcommand{\str}{$\sigma_{\rm tr}(T)$}
\newcommand{\mel}{$l_{\rm el}(T)$}
\newcommand{\pzero}{$\Phi_{0}$}

\title{Thermal conductivity and annealing effects in the iron-based superconductor FeSe$_{0.3}$Te$_{0.7}$}

\author{
	Masumi \textsc{Ohno}$^{1}$, 
	Takayuki \textsc{Kawamata}$^{1}$
	\thanks{E-mail address: tkawamata@teion.apph.tohoku.ac.jp}
	Takashi \textsc{Noji}$^{1}$, 
	Koki \textsc{Naruse}$^{1}$, 
	Yoshiharu \textsc{Matsuoka}$^{1}$, 
	Tadashi \textsc{Adachi}$^{1}$ 
	Terukazu \textsc{Nishizaki}$^{2}$ 
	Takahiko \textsc{Sasaki}$^{2}$ 
	and 
	Yoji \textsc{Koike}$^{1}$
}

\inst{
	$^{1}$Department of Applied Physics, Tohoku University, 6-6-05 Aoba, Aramaki, Aoba-ku, Sendai, Miyagi 980-8579, Japan
	$^{2}$Institute for Materials Research, Tohoku University, 2-1-1 Katahira, Aoba-ku, Sendai, Miyagi  980-8577, Japan
}

\abst{
Thermal conductivity measurements in magnetic fields have been carried out for FeSe$_{0.3}$Te$_{0.7}$ single crystals as-grown and annealed at 400 {\degree}C for 100 h in vacuum ($\sim$ 10$^{-4}$ Pa). It has been found that the thermal conductivity in the {\it ab}-plane, {\kab}, of the annealed crystal shows an enhancement at low temperatures just below the superconducting transition temperature, {\tc}, owing to the thermal conductivity due to quasiparticles, while {\kab} of the as-grown crystal does not.
This suggests that FeSe$_{1-x}$Te$_{x}$ is a strongly correlated electron system. It has also been found that both the degree of the enhancement of {\kab} just below {\tc} and the behavior of the suppression of {\kab} by the application of magnetic field for the annealed crystal depend on the thickness of the crystal.
These results indicate that bulk superconductivity is absent in the as-grown crystal, appears only in the surface area in the annealed-thick crystal and appears in the almost whole region in the annealed-thin crystal.
It has been concluded that the excess Fe included in the as-grown crystal is removed from the surface through the vacuum-annealing.
}

\vspace*{2em}

\kword{Fe-based superconductor, thermal conductivity, strongly correlated electron, annealing effect}

\begin{document}
\maketitle

\newpage

%***********************************************************************************************
\section{Introduction}
%***********************************************************************************************
The study of iron-based superconductors triggered by the discovery of superconductivity in the iron-pnictide LaFeAs(O,F) has been expanding and has remained active \cite{Hosono}, because there is a wide variety of crystal structures in iron-based superconductors as in the case of copper-based superconductors.
Among the iron-based superconductors, FeSe$_{1-x}$Te$_{x}$ has attracted great interest, because it has the simplest crystal structure composed of a stack of edge-sharing Fe(Se,Te)$_{4}$-tetrahedra layers which are similar to FeAs$_{4}$-tetrahedra layers in iron-pnictides \cite{FCHsu, KWYeh, MHFang}.
Single crystals of FeSe$_{1-x}$Te$_{x}$ have been grown in a range of $0.5 \le x \le 1.0$, but it has been reported by Sales {\it et al.} \cite{BCSales} that only single crystals with $x \sim 0.5$ exhibit bulk superconductivity.
Afterward, Noji {\it et al.} \cite{Noji2010,Noji2012} have revealed from magnetic-susceptibility and specific-heat measurements that single crystals with a wide range of $x =$ 0.5--0.9 exhibit bulk superconductivity through the annealing at 400 {\degree}C for 100--200 h in vacuum ($\sim$ 10$^{-4}$ Pa).
According to the EDX analysis in FeSe$_{0.39}$Te$_{0.61}$ by Taen {\it et al.} \cite{Taen}, the distribution of Se and Te in the crystal becomes homogeneous through the annealing.
Therefore, it has been believed that the homogeneous distribution of Se and Te through the annealing leads to the appearance of bulk superconductivity in a wide range of $x =$ 0.5--0.9, because the end member FeTe is not superconducting but develops an antiferromagnetic order at low temperatures below $\sim 67$ K \cite{SLi,Bao}.
However, it has been noticed that there is another factor more for the appearance of bulk superconductivity, because thick single-crystals do not exhibit bulk superconductivity in their whole regions through the vacuum-annealing.
Recently, it has been reported by Kawasaki {\it et al.} \cite{Kawasaki} that the annealing in oxygen for a short time operates to give rise to bulk superconductivity also and that this may be owing to the possible compensation for the positive charge of Fe excessively included in as-grown crystals of FeSe$_{1-x}$Te$_{x}$ by negative O$^{2-}$ ions introduced through the annealing.
On the other hand, it is known that the excess Fe suppresses the superconductivity in {\FeSeTe} \cite{TJLiu}.
Accordingly, it is possible that a small amount of oxygen introduced into an as-grown crystal through the annealing in vacuum ($\sim$ 10$^{-4}$ Pa) performed by Noji {\it et al.} \cite{Noji2010,Noji2012} eliminates the effect of the excess Fe, leading to the appearance of bulk superconductivity.

Thermal conductivity is a probe sensitive to the change of the electronic state in a superconductor, as used for the study of copper-based superconductors \cite{Kudo2004,Kawamata2006,Adachi2009}.
Based on the two-fluid model, electrons in the normal state change to be Cooper pairs and quasiparticles in the superconducting state. 
The Cooper pairs carry no heat and are not scattered by any other particles,
while quasiparticles carry heat and are scattered by the other particles except Cooper pairs. 
In the superconducting state, that is, the thermal conductivity due to electrons, $\kappa_{\rm el}$, decrease, because the number of electrons carrying heat decreases. 
Moreover, the thermal conductivity due to phonons, $\kappa_{\rm ph}$, increases, because the number of electrons scattering phonons decreases. 
The behavior of the thermal conductivity  of a conventional superconductor below the superconducting transition temperature, {\tc}, depends on the contributions of $\kappa_{\rm el}$ and $\kappa_{\rm ph}$.
It is well-known  that the thermal conductivity  of a pure metal decreases below {\tc} and that a small enhancement of the thermal conductivity is observed in an alloy where the contribution of $\kappa_{\rm el}$ is small below {\tc} \cite{Olsen1952}.
In strongly correlated electron systems, 
on the other hand, 
the contribution  of $\kappa_{\rm el}$ in the normal state to the thermal conductivity is not so large, 
because the electron-electron scattering is very strong. 
However, $\kappa_{\rm el}$ markedly increases in the superconducting  state owing to the decrease of the scattering between quasiparticles. 
%'à'¿'ë'ñ'±'Ì'Æ'«Cã‹L'Ì"dŽq'Ì"M"`"±''å'ÌŒø‰Ê'ƃtƒHƒmƒ"'Ì"M"`"±'Ì'‰Á'ÌŒø‰Ê'à‡'킳'Á'ÄŒ»'ê'éD
In copper-based superconductors with the $d$-wave symmetry, especially, the enhancement of $\kappa_{\rm el}$ below {\tc} is marked, for the number of quasiparticles is large owing to zero-gap nodes of the superconducting gap on the Fermi surfaces. 

In iron-based superconductors such as LaFeAs(O,F) \cite{ASSefat}, SmFeAs(O,F) \cite{Tropeano2008}, Ba(Fe,Co)$_{2}$As$_{2}$ \cite{Machida}, and (Ba,K)Fe$_{2}$As$_{2}$ \cite{Checkelsky}, an enhancement of the thermal conductivity in the {\it ab}-plane, {\kab}, at low temperatures just below {\tc} has been observed and the enhancement has been found to be suppressed by the application of magnetic field.
The electron-electron correlation in iron-based superconductors is not so strong as copper-based ones. 
Furthermore, the symmetry of iron-based superconductors is generally $s$-wave without node. 
Therefore, it is not so easy to clarify the origin of the enhancement of the thermal conductivity  below {\tc}.
Although it is possible that the enhancement of {\kab} just below {\tc} is owing to $\kappa_{\rm ph}$ \cite{Tropeano2008}, at present it is generally regarded as being owing to $\kappa_{\rm el}$ \cite{ASSefat,Machida,Checkelsky}.
That is, the electron-electron scattering is not so weak in the normal state of iron-based superconductors, 
leading to the reduction of the mean free path of electrons, {$l_{\rm el}$}.
Just below {\tc}, on the other hand, the number of quasiparticles carrying heat markedly decreases according to the two-fluid model 
as described above,
so that {$l_{\rm el}$} of quasiparticles is dramatically extended, leading to the enhancement of {\kab}.
The suppression of the enhancement of {\kab} by the application of magnetic field is interpreted as being caused by the decrease in {$l_{\rm el}$} due to the addition of the scattering of quasiparticles by vortices. 
In another interpretation, furthermore, the suppression is caused by the decrease in {$l_{\rm el}$} and/or the mean free path of phonon, {$l_{\rm ph}$}, due to the increase of the scattering of quasiparticles and/or phonons by quasiparticles on account of the reduction of the superconducting gap.

Thermal conductivity is a probe sensitive to the crystalline disorder also, 
because both {$l_{\rm el}$} and {$l_{\rm ph}$} are reduced by the crystalline disorder.
Therefore, both the inhomogeneity  of the distribution of Se and Te and the existence of the excess Fe in {\FeSeTe} are expected to influence the thermal conductivity.

As to the thermal conductivity of {\FeSeTe}, only Tropeano {\it et al.} \cite{Tropeano2010} have reported it using the polycrystals as far as we know, but no enhancement of the thermal conductivity just below {\tc} has been observed.
Neither has been obtained the information on the crystalline disorder, for the scattering of electrons and phonons by grain boundaries cannot be disregarded in polycrystals.
In this paper, we have measured {\kab} of as-grown and annealed single-crystals of {\Te} in magnetic-fields, in order to investigate vacuum-annealing effects on the enhancement of {\kab} just below {\tc} and the crystalline disorder.
Since it has been known that the vacuum-annealing effect on the superconductivity depends on the thickness of the crystal, two kinds of annealed single-crystals, namely, both thin and thick crystals have been used for the measurements.

%***********************************************************************************************
\section{Experimental}
%***********************************************************************************************
Single crystals of {\Te} were grown by the Bridgman method.
The details  are described in our previous paper \cite{Noji2010}.
As-grown crystals obtained thus were cloven in the {\abpl} to be thick ($\sim$ 1 mm in thickness along the {\it c}-axis) and thin ($\sim$ 0.1 mm in thickness) crystals and annealed at 400 {\degree}C for 100 h in vacuum ($\sim$ 10$^{-4}$ Pa) using an oil-diffusion pump.
The magnetic susceptibility, $\chi$, was measured using a superconducting quantum interference device (SQUID) magnetometer (Quantum Design, MPMS).
Measurements of {\kab} were carried out by the conventional steady-state method, where the temperature difference across the crystal was measured with two Cernox thermometers (LakeShore Cryotronics Inc., CX-1050-SD).
Since annealed-thin crystals were fragile, 
seven pieces of annealed-thin crystals
with the same size in the $ab$-plane were stacked 
and glued each other with GE7031 varnish and then used for the {\kab} measurements. 
It is noted that  the thermal conductivity of GE7031 varnish is negligibly small
Magnetic fields up to 14 T were applied parallel to the {\it c}-axis using a superconducting magnet.

%***********************************************************************************************
\section{Results and Discussion}
%***********************************************************************************************

Figure 1 shows the temperature dependence of $\chi$ on warming after zero field cooling for as-grown, annealed-thick and annealed-thin single-crystals of {\Te}.
For the as-grown crystal, the diamagnetic signal is small, though the electrical resistivity is zero below $\sim$ 14 K.
Therefore, it is concluded that not bulky but filamentary superconductivity takes place in the as-grown crystal.
Both annealed-thick and annealed-thin crystals, on the other hand, show large diamagnetic signals at low temperature below $\sim$ 14 K, indicating the appearance of bulk superconductivity.
However, this does not always imply that the whole regions in both annealed crystals are superconducting, because there remained a finite value of the electronic-specific-heat-coefficient at 0 K in the superconducting state of the thick crystal annealed at 400 {\degree}C for 100 h in vacuum ($\sim$ 10$^{-4}$ Pa) \cite{Noji2012}.

Figure 2 shows the temperature dependence of {\kab} in zero field for as-grown, annealed-thick and annealed-thin single-crystals of {\Te}.
It is found that {\kab}'s of all the crystals decrease monotonically with decreasing temperature down to {\tc}.
Below {\tc}, {\kab} of the as-grown crystal exhibits no change, while {\kab}'s of the annealed-thick and annealed-thin crystals show small and large enhancements, respectively, similar to {\kab} of the other iron-based superconductors \cite{ASSefat,Tropeano2008,Machida,Checkelsky}.
This is the first observation of the enhancement of {\kab} just below {\tc} in the {\FeSeTe} system.
Therefore, it is expected that the enhancement of {\kab} just below {\tc} is due to the enhancement of the contribution of quasiparticles to {\kab}, so that {\FeSeTe} is suggested to be a strongly correlated electron system also.
The difference of the degree of the enhancement just below {\tc} indicates that the superconducting region of the annealed-thin crystal is larger than that of the annealed-thick crystal, because the magnitude of the enhancement corresponds to the volume of the superconducting region in a crystal.
Taking into account that the diamagnetic signal of the annealed-thick crystal on warming after zero-field cooling shown in Fig. 1 is similar to that of the annealed-thin crystal, it is inferred that the annealed-thick crystal exhibits superconductivity only in the surface area, while the annealed-thin crystal does in the whole region.

Figures 3 (a), (b) and (c) show the temperature dependence of {\kab} in magnetic fields up to 14 T parallel to the {\it c}-axis on field cooling for as-grown, annealed-thick and annealed-thin single-crystals of {\Te}.
It is found that there is no change by the application of magnetic field in the as-grown crystal.
In both annealed-thick and annealed-thin crystals, on the other hand, the enhancement of {\kab} below {\tc} is suppressed by the application of magnetic field.
Figures 3 (d), (e) and (f) show the temperature dependence of {\kab}({\it H}) in magnetic field of {\it H} normalized by {\kab}(0) in zero field, {\kabHk}, for respective single-crystals.
In the as-grown crystal, certainly, there is no change by the application of magnetic field.
The slight change is due to the magnetic-field dependence of the Cernox thermometers.
In the annealed-thin crystal, {\kabHk} decreases at low temperatures below {\tc} and {\kabHk} decreases with increasing field.
In the annealed-thick crystal, on the other hand, {\kabHk} decreases below a different temperature dependent on the magnitude of magnetic field.
That is, the temperature below which {\kabHk} markedly decreases with decreasing temperature is lower than {\tc} in low fields and shifts to a high temperature up to {\tc} with increasing field.

Here, we consider the origin of the suppression of {\kab} below {\tc} by the application of magnetic field. 
If the suppression of the enhancement of {\kab} below {\tc} by the application of magnetic field is caused by the increase of quasiparticles  due to the reduction of the superconducting gap, 
the thermal conductivity is expected to start to be suppressed just below {\tc} in any magnetic fields. 
In the annealed thick-crystal, however, the temperature below which {\kab} is suppressed by the application of magnetic field is dependent on the magnetic-field strength. 
Therefore, the suppression of the enhancement of {\kab} below {\tc} by the application of magnetic field is expected to be due to the scattering of quasiparticles by vortices, as in the case of the other iron-based superconductors \cite{Machida,Checkelsky}.
These results also imply that the behavior of the vortex scattering in the superconducting region is different between annealed-thick and annealed-thin crystals, which may be understood based on the inference that the annealed-thick crystal exhibits superconductivity only in the surface area.
It is also found that the temperature dependence of {\kabHk} in the annealed-thin crystal shows an upturn at low temperatures below $\sim$ 6 K in high magnetic fields, but the origin is not clear.

In order to clarify the details of the suppression of the enhancement of {\kab} below {\tc} by the application of magnetic field, the data of {\kab} of the annealed-thin crystal were analyzed using the following equation based on the vortex-scattering model \cite{Machida,Checkelsky,Krishana}.
\begin{equation}
{\kappa}_{ab}(T,H) = \frac{{\kappa}_{\rm el}(T)}{1+{\alpha}(T)H} + {\kappa}_{\rm ph}(T),
\label{eq:vortex}
\end{equation}
where {\kel} and {\kph} are thermal conductivities due to electrons (quasiparticles) and phonons in zero field, respectively, and
\begin{equation}
{\alpha}(T) = l_{\rm el}(T){\sigma}_{\rm tr}(T)/{\Phi}_{0},
\label{eq:alpha}
\end{equation}
where {\str} is the cross section of the quasiparticle-vortex scattering and {\pzero} is the flux quantum.
In Eq. (\ref{eq:vortex}), it is assumed that only quasiparticles are scattered by vortices.
Figure 4 (a) shows the best-fit result in each magnetic field.
Figure 4 (b) displays values of the fitting parameters, {\kel} and {\kph}, obtained below {\tc}, together with experimental values of {\kab} in zero field, values of {\kel} above {\tc} estimated from the Wiedemann-Franz law and values of {\kph} above {\tc} calculated as ${\kappa}_{ab} - {\kappa}_{\rm el}(T)$.
It is found that values of {\kel} and {\kph} below {\tc} obtained from the fitting are smoothly connected with values of {\kel} and {\kph} above {\tc} estimated from the Wiedemann-Franz law and that {\kel} is dramatically enhanced below {\tc}.
This is very reasonable, because it is assumed in this analysis that only the enhancement of {\kel} below {\tc} is suppressed due to the scattering by vortices.
Assuming that only phonons are scattered by vortices, on the contrary, values of the fitting parameters, {\kel} and {\kph}, obtained below {\tc} are replaced with each other and the connection of these values with those above {\tc} estimated from the Wiedemann-Franz law becomes unnatural.
Accordingly, it is reasonable to understand that the enhancement of {\kab} below {\tc} is due to the enhancement of the contribution  of quasiparticles to {\kab} and that the suppression of the enhancement by the application of magnetic field is due to the decrease in $l_{\rm el}(T)$ caused by the quasiparticle-vortex scattering.

Figure 4 (c) shows values of the fitting parameter, {\aT}, obtained below {\tc}.
It is found that {\aT} is almost constant between {\tc} and $\sim$ 6 K and increases rapidly with decreasing temperature below $\sim$ 6 K, though it has been reported that {\aT} increases monotonically with decreasing temperature below {\tc} in the other iron-based superconductors \cite{Machida,Checkelsky}.
Since {\str} is roughly equal to the size of a vortex, namely, the coherence length, $\xi(T)$ \cite{Machida}, the almost constant behavior of {\aT} between {\tc} and $\sim$ 6 K is interpreted as being due to the balance of the increase in {\mel} and the decrease in $\xi(T)$ with decreasing temperature.
Moreover, it is understood that the increase in {\mel} with decreasing temperature just below {\tc} is not so marked as in the other iron-based superconductors.
The rapid increase in {\aT} with decreasing temperature below $\sim$ 6 K is interpreted as being due to the rapid increase in {\mel}.
These results may be in correspondence to the result in FeSe$_{0.4}$Te$_{0.6}$ obtained from the microwave conductivity by Takahashi {\it et al.}\cite{Takahashi} that the superconducting state is in the dirty limit just below {\tc} and changes to be in the clean limit at low temperatures.
Putting ${\xi}(2{\rm K})$ at $\sim$ 15 {\AA}, which has been estimated from the value of the  upper critical field \cite{Takahashi,HLei,Klein,Imai}, $l_{\rm el}(2{\rm K})$ is estimated as $\sim$ 2800 {\AA}.
This value is of the same order as that in Ba(Fe,Co)$_{2}$As$_{2}$ \cite{Machida}.
The present estimated values of $l_{\rm el}(T)$ seem to be too large considering that iron-based superconductors are rather regarded as strongly correlated electron systems. 
This is presumably due to the assumption that the value of {\str} is equal to that of $\xi(T)$. 

Here, we discuss the superconducting region in as-grown, annealed-thick and annealed-thin single-crystals of {\Te}.
In the as-grown crystal, the superconductivity is not bulky but filamentary.
Therefore, it is reasonable that there is neither enhancement of {\kab} below {\tc} nor magnetic-field effect, because there is neither marked decrease of quasiparticles below {\tc} nor generation of vortices.
In the annealed-thin crystal, on the other hand, it is no doubt that bulk superconductivity appears in the almost whole region, because the temperature and magnetic-field dependences of {\kab} are well explained as described above.
In the annealed-thick crystal, since there remains a normal region at 0 K in the superconducting state of the bulk according to the specific-heat measurements \cite{Noji2012}, it seems that the bulk superconductivity appears only in the surface area and that the inside exhibits superconductivity as filamentary as in the as-grown crystal.
Therefore, it is reasonable that the enhancement of {\kab} below {\tc} is not so marked in the annealed-thick crystal as in the annealed-thin crystal where bulk superconductivity appears in the almost whole region.
Moreover, the magnetic-field dependence of the temperature below which {\kabHk} markedly decreases with decreasing temperature is well explained as follows.
Considering that the suppression of {\kab} just below {\tc} appears in magnetic fields above $\sim$~7~T, the size of the superconducting region in the surface area is inferred to be comparable with the distance between vortices in $\sim$~7~T just below {\tc} in the annealed-thick crystal.
That is, on cooling in a low field, the number of vortices is small just below {\tc}, because the superconducting region is narrow and moreover the distance between vortices is long.
Therefore, the enhancement of {\kab} is not suppressed so much.
With decreasing temperature, the superconducting region gradually extends, so that the number of vortices increases, leading to the suppression of the enhancement of {\kab}.
On cooling in a high field, on the other hand, the number of vortices is not small even just below {\tc}, because the distance between vortices becomes short.
Therefore, the temperature below which {\kabHk} marked decreases with decreasing temperature increases with increasing field.

Finally, we discuss the vacuum-annealing effect on the superconductivity.
As described in I, taking into account both the dependence on the thickness of a single crystal of the vacuum-annealing effect and the existence of the excess Fe in an as-grown crystal suppressing the superconductivity \cite{TJLiu}, it is reasonable to understand that the excess Fe is removed from the surface of the as-grown crystal through the vacuum-annealing.
That is, the excess Fe is inferred to be present homogeneously in the as-grown crystal.
In the annealed-thick crystal, it seems that there remains the excess Fe in the inside of the crystal where the superconductivity is suppressed.
In the annealed-thin crystal, on the other hand, it seems that there remains no excess Fe, so that bulk superconductivity appears in the almost whole region.
Furthermore, it is inferred that the excess Fe reacts with oxygen included in vacuum ($\sim$ 10$^{-4}$ Pa) in the surface area to form iron-oxides and be excluded from the surface, so that the excess Fe in the inside of the crystal diffuses to the surface area and reacts with oxygen.
This inference is consistent with the result that {\kab} in the annealed-thin crystal is larger than {\kab} in as-grown and annealed-thick crystals, because the excess Fe randomly included in as-grown and annealed-thick crystals is expected to scatter heat carriers.
In order to compensate the effect of the excess Fe, the intercalation of oxygen has been suggested to occur by Kawasaki {\it et al.} \cite{Kawasaki}.
However, this seems unreasonable, because the intercalation of oxygen must introduce crystalline disorder locally and decrease {\kab} in the annealed crystals more than in the as-grown crystal.

%***********************************************************************************************
\section{Summary}
%***********************************************************************************************
We have measured {\kab} in magnetic fields of {\Te} single crystals as-grown and annealed at 400 {\degree}C for 100 h in vacuum ($\sim$ 10$^{-4}$ Pa).
At low temperatures just below {\tc}, it has been found that {\kab} of the as-grown crystal is not enhanced, while {\kab}'s of annealed-thick and annealed-thin crystals show small and large enhancements corresponding to the region of bulk superconductivity, respectively.
In the annealed-thin crystal, the enhancement of {\kab} is suppressed by the application of magnetic field at low temperatures just below {\tc}.
In the annealed-thick crystal, on the other hand, the temperature below which the enhancement is suppressed by the application of magnetic field is dependent on the magnetic field and decreases with decreasing magnetic-field.

From the analysis based on the vortex-scattering model, it has been concluded that the enhancement of {\kab} just below {\tc} is due to the enhancement of the contribution of quasiparticles to {\kab} and that the suppression of the enhancement by the application of magnetic field is due to the decrease in {\it l}$_{\rm el}$ caused by the quasiparticle-vortex scattering.
Accordingly, {\FeSeTe} is concluded to be a strongly correlated electron system as the other iron-based superconductors

As for the superconducting region in these single-crystals, the superconductivity in the as-grown crystal is not bulky but filamentary.
It has been concluded that bulk superconductivity appears only in the surface area in the annealed-thick crystal, while it appears in the almost whole region in the annealed-thin crystal.
Moreover, it has been concluded that the excess Fe included in the as-grown crystal suppressing the superconductivity is removed from the surface through the vacuum-annealing as a result of the reaction in the surface area between the excess Fe and oxygen included in vacuum ($\sim~10^{-4}$ Pa), so that the inside of the annealed-thick crystal does not exhibit bulk superconductivity on account of the residual of the excess Fe.

%***********************************************************************************************
\section{Acknowledgments}
%***********************************************************************************************
The thermal conductivity measures were performed at the High Field Laboratory for Superconducting Materials, Institute for Materials Research, Tohoku University.

%***********************************************************************************************

\newpage

%***********************************************************************************************
\section*{Figure captions}
%***********************************************************************************************
\begin{description}
\item{Fig. 1.}
Temperature dependence of the magnetic susceptibility, $\chi$, on warming after zero-field cooling for as-grown, annealed-thick and annealed-thin single-crystals of {\Te}.

\item{Fig. 2.}
Temperature dependence of the thermal conductivity in the {\it ab}-plane, {\kab}, in zero field for as-grown, annealed-thick and annealed-thin single-crystals of {\Te}.

\item{Fig. 3.}
(a)--(c) Temperature dependence of the thermal conductivity in the {\it ab}-plane, {\kab}, in various magnetic fields up to 14 T parallel to the {\it c}-axis on field cooling for as-grown, annealed-thick and annealed-thin single-crystals of {\Te}.
(d)--(f) Temperature dependence of {\kab}($H$) in magnetic field of $H$ normalized by {\kab}(0) in zero field, {\kabHk}, for respective single-crystals.

\item{Fig. 4.} (a) Magnetic-field dependence of the normalized thermal conductivity, {\kabHk}, at various temperatures.
Solid lines indicate the best-fit results using Eq. (\ref{eq:vortex}) based on the vortex-scattering model.
(b) Temperature dependence of the fitting parameters, {\kel} (filled circle) and {\kph} (filled triangle), in Eq.(\ref{eq:vortex}) obtained below {\tc}.
Both experimental values of {\kab} in zero field (filled square), values of {\kel} above {\tc} estimated from the Wiedemann-Frans law (open circle) and values of {\kph} above {\tc} calculated as ${\kappa}_{ab} - {\kappa}_{\rm el}(T)$ (open triangle) are shown also.
(c) Temperature dependence of the fitting parameter, {\aT}, in Eq. (\ref{eq:vortex}) obtained below {\tc}.
(d) Temperature dependence of the mean free path of quasiparticles, {\mel}, obtained using Eq.(\ref{eq:alpha}).

\end{description}

\begin{figure}[p]
	\begin{center}
		\includegraphics[scale=1]{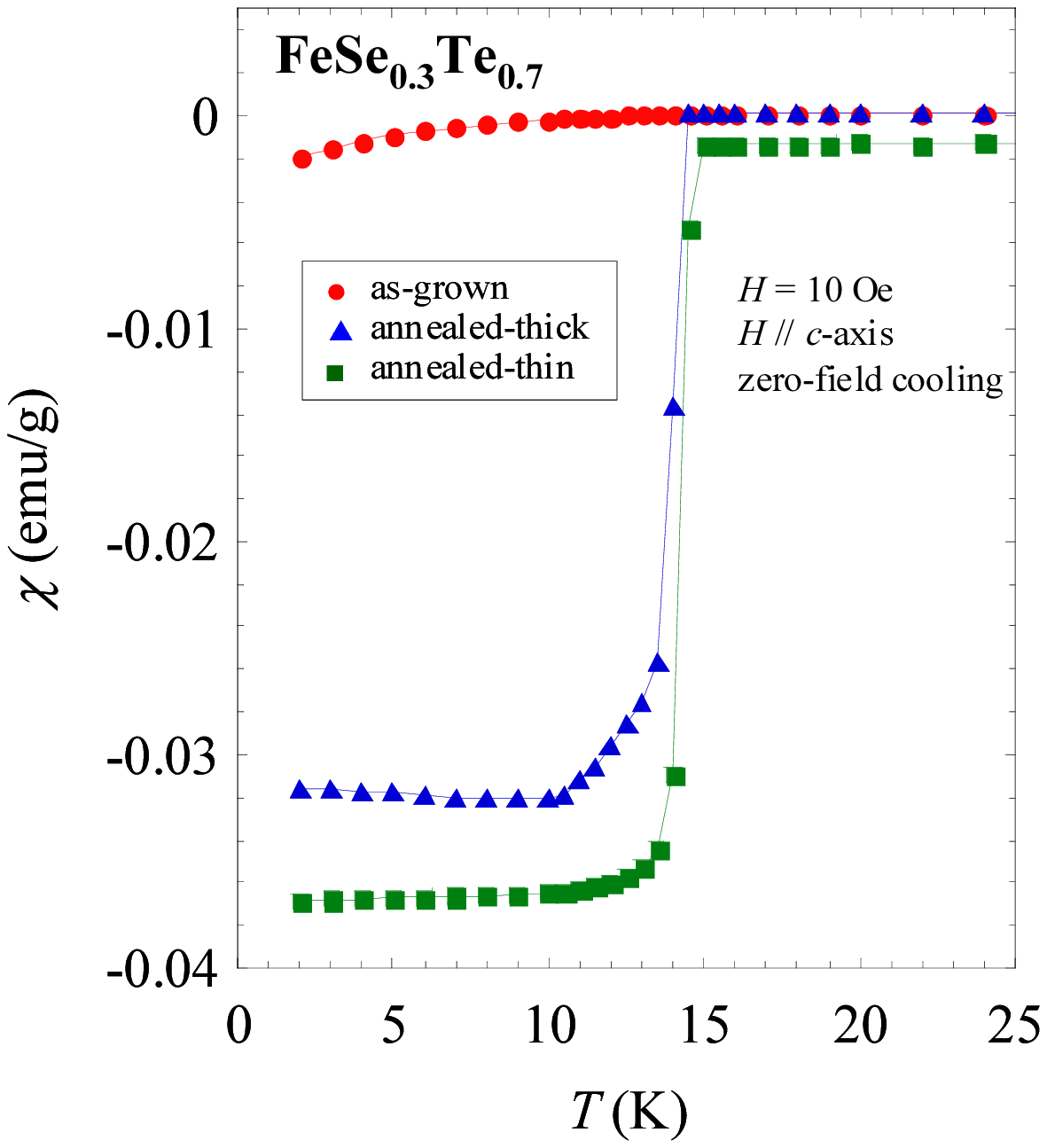}
		\caption{}
	\end{center}
\end{figure}

\begin{figure}[p]
	\begin{center}
		\includegraphics[scale=1]{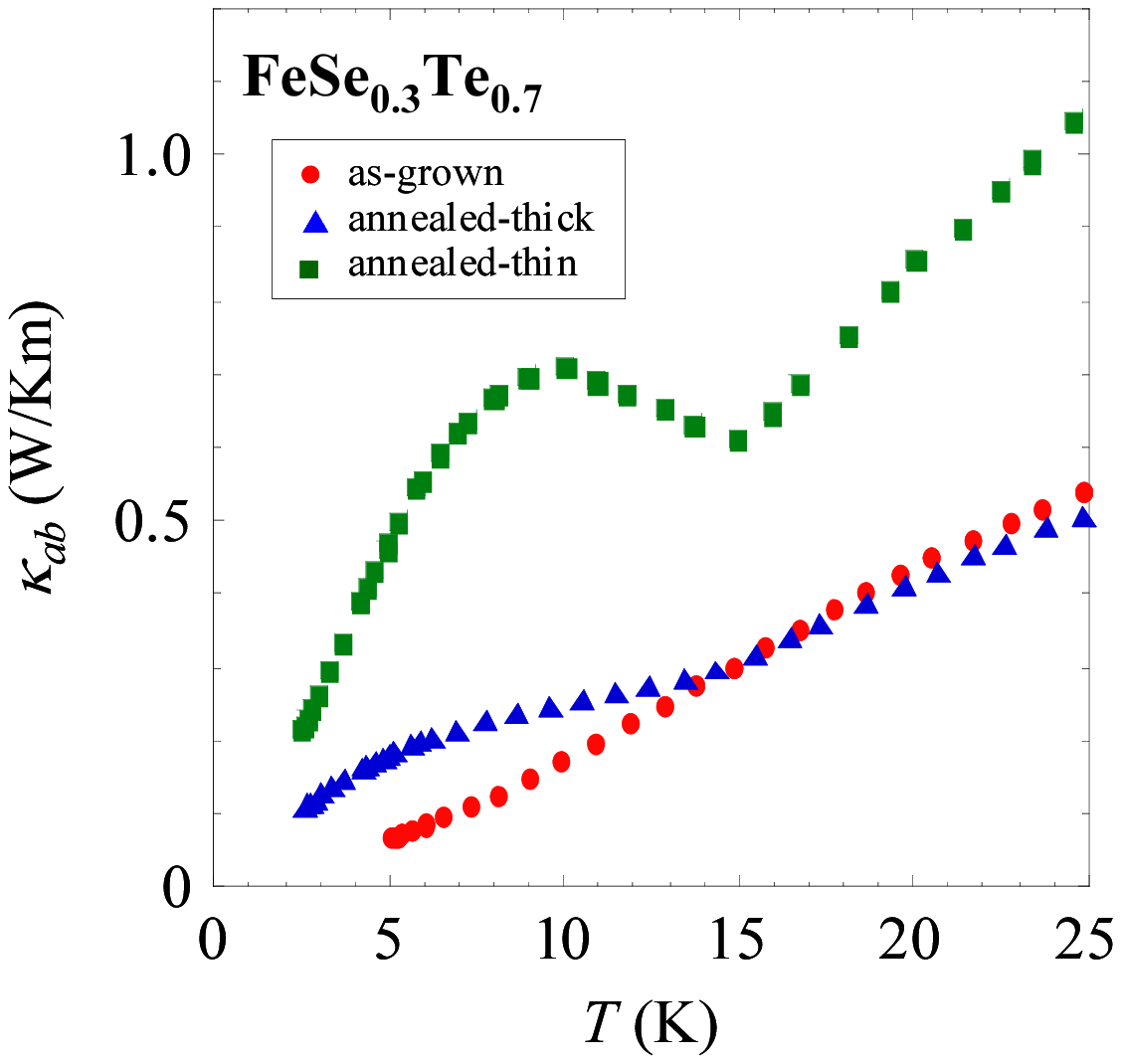}
		\caption{}
	\end{center}
\end{figure}

\begin{figure}[p]
	\begin{center}
		\includegraphics[scale=0.75]{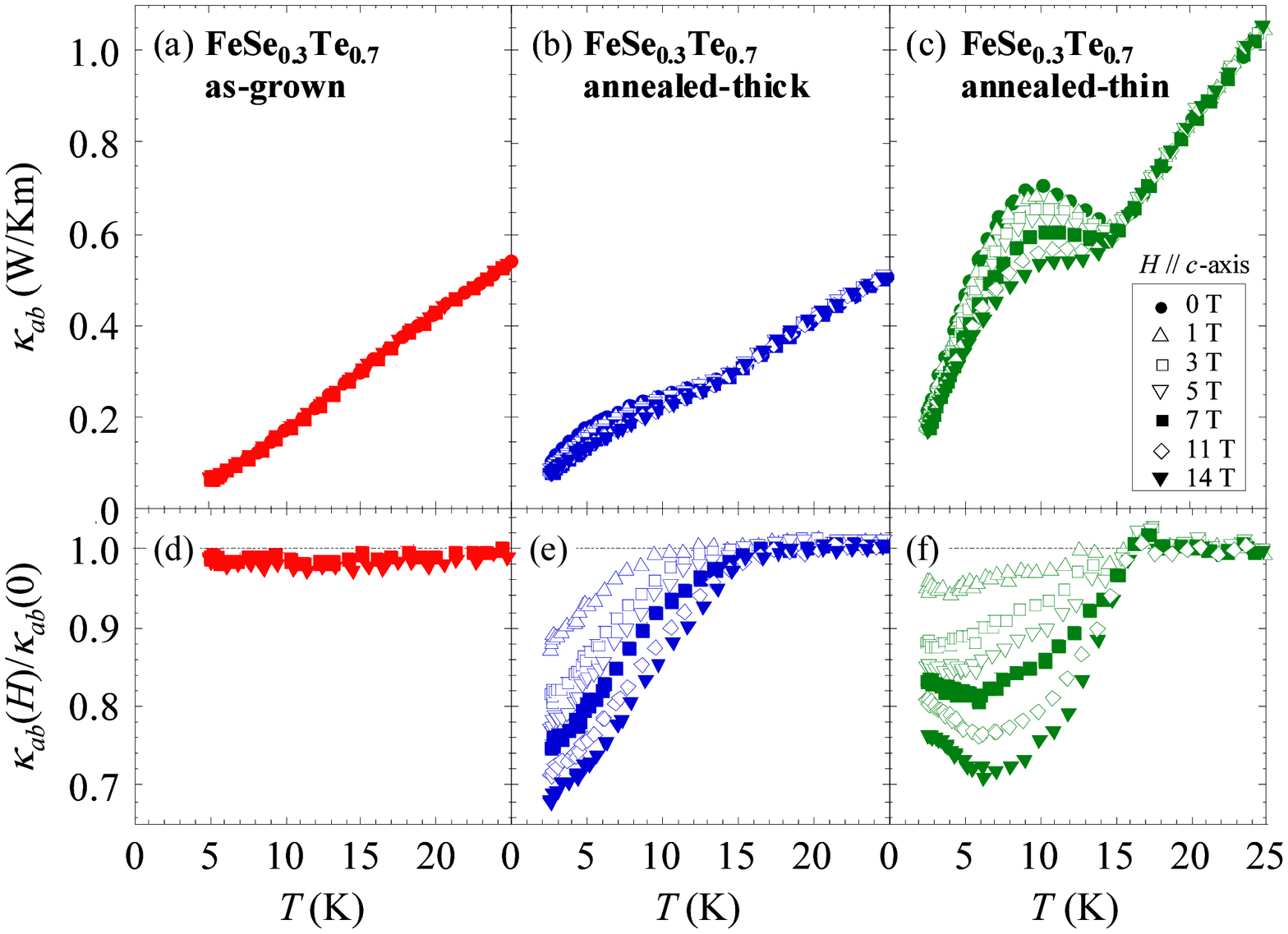}
		\caption{}
	\end{center}
\end{figure}

\begin{figure}[p]
	\begin{center}
		\includegraphics[scale=0.7]{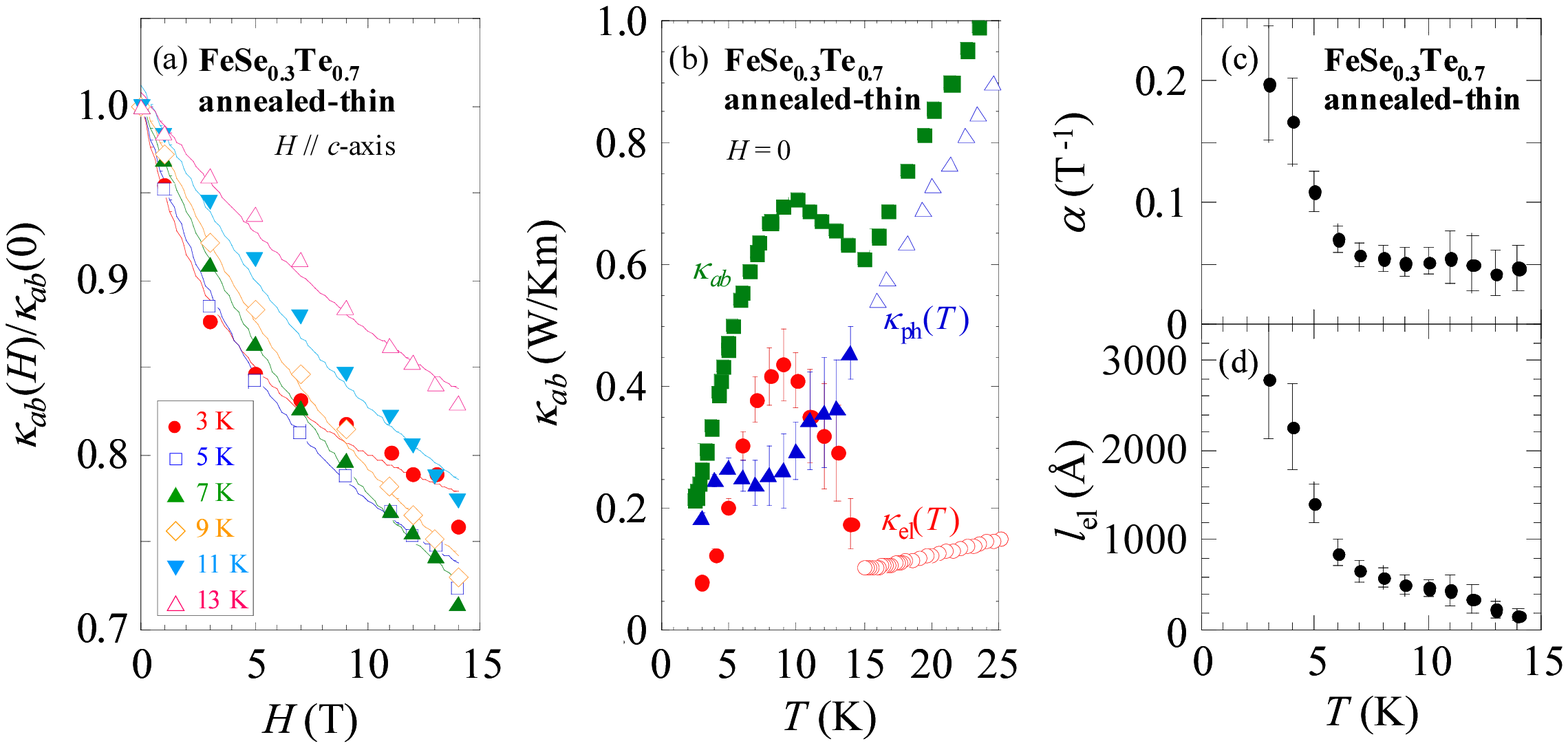}
		\caption{}
	\end{center}
\end{figure}


\begin{thebibliography}{99}
\bibitem{Hosono}
Y. Kamihara, T. Watanabe, M. Hirano, and H. Hosono:
 J. Am. Chem. Soc. {\bf 130} (2008) 3296.

\bibitem{FCHsu}
F.-C. Hsu, J.-Y. Luo, K.-W. Yeh, T.-K. Chen, T.-W. Huang, P. M. Wu, Y.-C. Lee, Y.-L. Huang, Y.-Y. Chu, D.-C. Yan, and M.-K. Wu:
 Proc. Natl. Acad. Sci. U.S.A. {\bf 105} (2008) 14262.

\bibitem{KWYeh}
K.-W. Yeh, T.-W. Huang, Y.-L. Huang, T.-K. Chen, F.-C. Hsu, P. M. Wu, Y.-C. Lee, Y.-Y. Chu, C.-L. Chen, J.-Y. Luo, D.-C. Yan, and M.-K. Wu:
 Europhys. Lett. {\bf 84} (2008) 37002.

\bibitem{MHFang}
M. H. Fang, H. M. Pham, B. Qian, T. J. Liu, E. K. Vehstedt, Y. Liu, L. Spinu, and Z. Q. Mao:
 Phys. Rev. B {\bf 78} (2008) 224503.

\bibitem{BCSales}
B. C. Sales, A. S. Sefat, M. A. McGuire, R. Y. Jin, D. Mandrus, and Y. Mozharivskyj:
 Phys. Rev. B {\bf 79} (2009) 094521.

\bibitem{Noji2010}
T. Noji, T. Suzuki, H. Abe, T. Adachi, M. Kato, and Y. Koike:
 J. Phys. Soc. Jpn. {\bf 79} (2010) 084711.

\bibitem{Noji2012}
T. Noji, M. Imaizumi, T. Suzuki, T. Adachi, M. Kato, and Y. Koike:
 J. Phys. Soc. Jpn. {\bf 81} (2012) 054708.

\bibitem{Taen}
T. Taen, Y. Tsuchiya, Y. Nakajima, and T. Tamegai:
 Phys. Rev. B {\bf 80} (2009) 092502.

\bibitem{SLi}
S. Li, C. de la Cruz, Q. Huang, Y. Chen, J. W. Lynn, J. Hu, Y.-L. Huang, F.-C. Hsu, K.-W. Yeh, M.-K. Wu, and P. Dai:
 Phys. Rev. B {\bf 79} (2009) 054503.

\bibitem{Bao}
W. Bao, Y. Qiu, Q. Huang, M. A. Green, P. Zajdel, M. R. Fitzsimmons, M. Zhernenkov, S. Chang, M. Fang, B. Qian, E. K. Vehstedt, J. Yang, H. M. Pham, L. Spinu, and Z. Q. Mao:
 Phys. Rev. Lett. {\bf 102} (2009) 247001.

\bibitem{Kawasaki}
Y. Kawasaki, K. Deguchi, S. Demura, T. Watanabe, H. Okazaki, T. Ozaki, T. Yamaguchi, H. Takeya, and Y. Takano:
 Solid State Commun. {\bf 152} (2012) 1135.

\bibitem{TJLiu}
T. J. Liu, X. Ke, B. Qian, J. Hu, D. Fobes, E. K. Vehstedt, H. Pham, J. H. Yang, M. H. Fang, L. Spinu, P. Schiffer, Y. Liu, and Z. Q. Mao:
 Phys. Rev. B {\bf 80} (2009) 174509.

\bibitem{Kudo2004}
K. Kudo, M. Yamazaki, T. Kawamata, T. Adachi, T. Noji, Y.Koike, T. Nishizaki, and N. Kobayashi:
 Phys. Rev. B {\bf 70} (2004) 014503.

\bibitem{Kawamata2006}
T. Kawamata, N. Takahashi, M. Yamazaki, T. Adachi, T. Manabe, T. Noji, Y. Koike, K. Kudo, and N. Kobayashi:
 AIP Conf. proc. {\bf 850} (2006) 431.

\bibitem{Adachi2009}
T. Adachi, S. M. Haidar, T. Kawamata, N. Sugawara, N. Kaneko, M. Uesaka, H. Sato, Y. Tanabe, T. Noji, K. Kudo, N. Kobayashi, and Y. Koike:
 J. Phys.: Conf. Ser. {\bf 150} (2009) 052115.

\bibitem{Olsen1952}
J. L. Olsen: Proc. Phys. Soc. {\bf A65} (1952) 518.

\bibitem{ASSefat}
A. S. Sefat, M. A. McGuire, B. C. Sales, R. Jin, J. Y. Howe, and D. Mandrus:
 Phys. Rev. B {\bf 77} (2008) 174503.

\bibitem{Tropeano2008}
M. Tropeano, A. Martinelli, A. Palenzona, E. Bellingeri, E. Galleani d'Agliano, T. D. Nguyen, M. Affronte, and M. Putti:
 Phys. Rev. B {\bf 78} (2008) 094518.

\bibitem{Machida}
Y. Machida, K. Tomokuni, T. Isono, K. Izawa, Y. Nakajima, and T. Tamegai:
 J. Phys. Soc. Jpn. {\bf 78} (2009) 073705.

\bibitem{Checkelsky}
J. G. Checkelsky, R. Thomale, Lu Li, G. F. Chen, J. L. Luo, N. L. Wang, and N. P. Ong:
 Phys. Rev. B {\bf 86} (2012) 180502(R).

\bibitem{Tropeano2010}
M. Tropeano, I. Pallecchi, M. R. Cimberle, C. Ferdeghini, G. Lamura, M. Vignolo, A. Martinelli, A. Palenzona, and M. Putti:
 Supercond. Sci. Technol. {\bf 23} (2010) 054001.

\bibitem{Krishana}
K. Krishana, N. P. Ong, Y. Zhang, Z. A. Xu, R. Gagnon, and L. Taillefer:
 Phys. Rev. Lett. {\bf 82} (1999) 5108.

\bibitem{Takahashi}
H. Takahashi, Y. Imai, S. Komiya, I. Tsukada, and A. Maeda:
 Phys. Rev. B {\bf 84} (2011) 132503.

\bibitem{HLei}
H. Lei, R. Hu, E. S. Choi, J. B. Warren, and C. Petrovic:
 Phys. Rev. B {\bf 81} (2010) 094518.

\bibitem{Klein}
T. Klein, D. Braithwaite, A. Demuer, W. Knafo, G. Lapertot, C. Marcenat, P. Rodi\'{e}re, I. Sheikin, P. Strobel, A. Sulpice, and P. Toulemonde:
 Phys. Rev. B {\bf 82} (2010) 184506.

\bibitem{Imai}
Y. Imai, R. Tanaka, T. Akiike, M. Hanawa, I. Tsukada, and A. Maeda:
 Jpn. J. Appl. Phys. {\bf 49} (2010) 023101.

\end{thebibliography}
\end{document}